\documentclass[preprintnumbers，floatfix, letterpaper, twocolumn, superscriptaddress,nofootinbib]{revtex4-2}
\usepackage{graphicx}
\usepackage{amsmath}
\usepackage{amssymb}
\usepackage{subfigure}
\usepackage{hyperref}
\usepackage{url}
\usepackage{xcolor}
\usepackage{color}
\usepackage{mathrsfs}
\usepackage{calrsfs}
\usepackage{amsfonts}
\usepackage{tabularx}
\usepackage{eucal}
\usepackage{latexsym}
\usepackage{ragged2e}
\usepackage{epsfig}
\usepackage{textcomp}
\usepackage{float}
\usepackage{wasysym}

\usepackage{caption}
\DeclareCaptionJustification{justified}{\leftskip=0pt \rightskip=0pt \parfillskip=0pt plus 1fil}
\definecolor{vividviolet}{rgb}{0.62, 0.0, 1.0}
\definecolor{amaranth}{rgb}{0.9, 0.17, 0.31}
\definecolor{palatinateblue}{rgb}{0.15, 0.23, 0.89}
\definecolor{brightpink}{rgb}{1.0, 0.0, 0.5}
\definecolor{cornflowerblue}{rgb}{0.39, 0.58, 0.93}
\definecolor{deepcarminepink}{rgb}{0.94, 0.19, 0.22}
\definecolor{radicalred}{rgb}{1.0, 0.21, 0.37}

\hypersetup{ linktoc=all,
	colorlinks, linkcolor={palatinateblue},
	citecolor={brightpink}, urlcolor={amaranth}
}

\graphicspath{{Images/}}

\begin{document}
\title{Upper Bound of Barrow Entropy Index From Black Hole Fragmentation}

\author{Jiayi \surname{Xia}}
\affiliation{Center for Gravitation and Cosmology, College of Physical Science and Technology, Yangzhou University, \\180 Siwangting Road, Yangzhou City, Jiangsu Province  225002, China}
	
\author{Yen Chin \surname{Ong}}
\email{ycong@yzu.edu.cn}
\affiliation{Center for Gravitation and Cosmology, College of Physical Science and Technology, Yangzhou University, \\180 Siwangting Road, Yangzhou City, Jiangsu Province  225002, China}
\affiliation{Shanghai Frontier Science Center for Gravitational Wave Detection, School of Aeronautics and Astronautics, Shanghai Jiao Tong University, Shanghai 200240, China}

\begin{abstract}
Both classical and quantum arguments suggest that if Barrow entropy is correct, its index $\delta$ must be energy dependent, which would affect the very early universe. Based on thermodynamic stability that sufficiently large black holes should not fragment, we argue that Barrow entropy correction must be small except possibly at the Planckian regime. Furthermore, the fact that a solar mass black hole does not fragment implies an upper bound $\delta \lesssim O(10^{-3})$, which surprisingly lies in the same range as the bound obtained from some cosmological considerations assuming fixed $\delta$. This indicates that allowing $\delta$ to run does not raise its allowed value. We briefly comment on the case of Kaniadakis entropy.
\end{abstract} 

\maketitle

\section{Introduction: How Big Can Barrow Entropy Correction Be?}
Barrow proposed that black hole horizons are ``fractalized'' by quantum gravitational (QG) effects into a ``sphereflake'' \cite{2004.09444} . The entropy of a black hole is 
\begin{equation}\label{Barrow}
S_B={k_\text{B}} \left(\frac{A}{4\ell_p^2}\right)^{1+\frac{\delta}{2}},
\end{equation}
where $\ell_p^2$ is Planck area and $k_\text{B}$ the Boltzmann constant\footnote{Hereinafter, we set $c=G=\hbar=k_\text{B}=1$.}. We refer to $\delta \in [0,1]$ as the Barrow Entropy Index (BEI). This concept also generalizes to the apparent horizon of a cosmological spacetime.

In \cite{2205.09311}, BEI is proposed to be energy-scale dependent due to the following result (see also \cite{2203.12010,2205.04138,2112.10159}):
the small $\delta$ expansion of Barrow entropy is
\begin{equation}\label{BEI}
S_B = \frac{A}{4} + \frac{A}{8} \ln\left(\frac{A}{4}\right)\delta +O(\delta^2).
\end{equation}
If $\delta$ is fixed, then for sufficiently large black holes, the second ``correction'' term will dominate the first term no matter how small $\delta$ is\footnote{This argument should also be applicable to the Tsallis entropy \cite{tsallis} which can be similarly expanded \cite{2109.05315}, though BEI is explicitly motivated as being a QG effect while Tsallis entropy is a statistical mechanical generalization.}. This is not reasonable because we expect a large black hole to receive little QG correction, and its entropy should satisfy the standard Bekenstein-Hawking area law. Such a deviation from the area law is in tension with holography as well. Eq.(\ref{BEI}) also disagrees with the results from various quantum gravity theories/models\footnote{Barrow \cite{2004.09444} also motivated his proposal with the generalized uncertainty principle (GUP), however GUP too gives $\ln A$ as the first order QG correction term instead of $A\ln A$; see \cite{0106080}. Barrow's own argument therefore supports a running BEI.}, which require that the leading correction must go like $\ln A$. Both classical and QG arguments thus imply $\delta$ should run, and at least scale like $\delta \sim 1/(A \ln A)$ for large $A$ so that it is sub-leading to both the area law and the standard QG corrections. 

{We remark that $\delta \sim 1/(A \ln A)$ as opposed to $\delta \sim 1/A$ is essentially a choice. Following \cite{2205.09311}, the idea is that Barrow correction could be a substantially different quantum correction independent of the standard QG correction, as it is more interesting to explore possible new physics that could arise from a novel subdominant correction to the usual logarithmic correction. In fact, Barrow correction had been imposed on top of the standard QG correction in \cite{MSalehi}. 
Nevertheless, both choices $\delta \sim 1/(A \ln A)$ and $\delta \sim 1/A$ are distinct from the standard QG one, which can be seen by considering small BEI expansion.
The choice $\delta \sim 1/(A \ln A)$ yields higher order correction terms which are inverse powers of the area, i.e., of the same form as the standard QG version, whereas the choice $\delta \sim 1/A$, while providing a logarithmic term, also gives rise to higher order correction of the form $(\ln A)^k/A^{k-1}$. In principle this is possible, and would provide a larger correction to small quantum black holes. The physical implications deserve a further investigation. 

Regardless, this means that astrophysical black holes have extremely small value of $\delta$. For a solar mass black hole, $M = M_\text{\astrosun} \sim 10^{38}$, so $\delta \sim 10^{-80}$ if $A \sim 1/A \ln A$.
If we choose $\delta \sim 1/A$ instead, we get $\delta \sim 10^{-78}$, still a very tiny number. Even at the Planck scale\footnote{Of course, too close to the Planck scale, higher order terms in Eq.(\ref{BEI}) as well as other QG effects would also start to become important.}, $\delta  \sim 0.005$ and $0.02$ for $\delta \sim 1/A \ln A$ and $\delta \sim 1/A$, respectively.}

Regardless, physically we expect $\delta$ to be essentially zero for large black holes and increases as a black hole size decreases towards the Planck scale. The horizon of an old black hole -- to use the term in \cite{2006.09385} -- would thus ``wrinklify''  as the black hole loses mass under Hawking evaporation. A natural and important question is: how large can the BEI be?
In \cite{2205.09311}, early universe cosmology was analyzed by assuming that $\delta$ decreases sufficiently (in fact, exponentially) fast as the universe expands. As already mentioned in that work, clearly infinitely many functions would satisfy $\delta \sim 1/(A \ln A)$ or $\delta \sim 1/A$, and there is no additional equation that governs the dynamic of $\delta$; an exponentially decreasing function was chosen for its simplicity. Thus there is a huge amount of freedom for how the function $\delta(A)$, or more precisely, $\delta(M)$ as a function of black hole mass, might actually look like. Can $\delta$ be quite large, i.e., close to unity, until a certain critical $M$ is reached, whereby its value drops suddenly?

Thinking along this line, one might claim that perhaps there exists a mass scale ${M}^* \gg 1$ such that $M \geqslant M^*$ is considered as large. This is equivalent to say that the usual area law breaks down for black hole whose mass is smaller than $M^*$. This is unlikely, but the point we wish to make is that even if we consider such a scenario, the value of $\delta$ remains small. 
The smallest black hole ever observed is of the order of a solar mass. If $M^*=M_\text{\astrosun}$, by definition black holes with mass $M \geqslant M^*$ satisfies $\delta \sim 1/(A \ln A)$, whereas it is possible that a black hole whose mass is smaller than $M^*$ to admit a much larger value for $\delta$. We can construct such a $\delta$ smoothly via bump functions with arbitrarily small transition width $\epsilon$, so that $\delta \gg 1/(A \ln A)$ or $\delta \sim 1/A$ for $M < M^*-\epsilon$. We want to investigate how large can $\delta$ be under such a scenario.  {Note that our subsequent argument does not depend on the choice of either $\delta \sim 1/(A \ln A)$ or $\delta \sim 1/A$}. 

\section{Bounding Barrow Entropy Index with Black Hole Fragmentation}
We shall show that the black hole with mass ${M}^*$ cannot be stable -- it will fragment into two black holes with a slightly smaller mass.
To see this, we compare the initial entropy 
\begin{equation}
S_B(M^*) = (4\pi)^{1+\frac{\delta_i}{2}} (M^*)^{2+\delta_i},
\end{equation}
with the entropy of the final configurations whose BEI is denoted by $\delta_f$:
\begin{equation}
2 S_B\left({M^*}/{2}\right) = 2\times (4\pi)^{1+\frac{\delta_f}{2}} \left(\frac{M^*}{2}\right)^{2+\delta_f}.
\end{equation}
We can neglect the initial BEI $\delta_i \sim 10^{-80}$ for $M^*=M_\text{\astrosun}$. Fragmentation can happen if the final configuration is entropically favored, i.e., if
\begin{equation}
{4\pi (M^*)^2}<2\pi^{1+\frac{\delta_f}{2}}(M^*)^{2+\delta_f}.
\end{equation}
For $M^*=M_\text{\astrosun}$, fragmentation happens if $\delta_f \gtrsim 0.00787$. That is, the fact that we observed black holes with $M \sim M_\text{\astrosun}$ implies that $\delta \lesssim 0.00787$
(We can also consider fragmentation into more than two black holes, but this does not give a better bound.) If we ever observe an even smaller black hole, the upper bound for $\delta$ can be increased. But even if 
$M^*=10^2$, just 100 times greater than Planck mass, we have $\delta \lesssim 0.13388$. Thus introducing the cutoff mass $M^*$ also does not permit $\delta$ to be larger than $O(0.1)$. Again, as per Footnote 4, we emphasize that other effects may come into play before the Planck scale. For example, if spacetime is non-commutative, the non-commutative scale can be much larger than the Planck scale \cite{1901.01613}: the upper bound is some 10 orders of magnitude larger than Planck length -- by considering gamma ray burst data (since non-commutativity can influence the group velocity of massless particle wave packets). So if we stay in the regime $M^* \gg 1$, say $M^*_{\text{min}}=10^6$ for definiteness, then $\delta \lesssim 0.048176$.

\section{Discussion}
Two remarks are in order. Firstly: the fact that black holes can fragment is intimately related to the deviation from the standard area law. To see that, consider any modification of entropy of the form
\begin{equation}
S=\frac{A}{4}\left[1+\varepsilon F(A) + O(\varepsilon^2)\right]=4\pi M^2\left[1+\varepsilon F(M)\right],
\end{equation}
expanded as a series of the small deviation parameter $\varepsilon$, which itself is running, i.e., a function of the mass just like $F(M)$.
If $\varepsilon_i \approx 0$, we see that the final entropy,
\begin{equation}
S_f = 2 \times 4\pi \left(\frac{M}{2}\right)^2 \left[1+\varepsilon_f F\left(\frac{M}{2}\right)\right], 
\end{equation}
can only be larger than the initial entropy $S_i \sim 4\pi M^2$, if $\varepsilon_f F(M/2) > 1$, i.e., the ``correction'' term dominates over the original area law term.

Lastly, the bound we obtained from the fact that solar mass black holes are stable, $\delta \lesssim O(10^{-3})$, rather surprisingly falls in the same range as the bound obtained from various cosmological considerations assuming fixed BEI \cite{2005.10302,2010.00986,2108.10998,2203.12010}. In other words, a running BEI does not allow larger values of $\delta$ simply from thermodynamic stability. 

The argument presented in this short note is also applicable to other entropy proposals. As commented in Footnote 2, Tsallis entropy has almost the same form so the result can be readily transformed over. One can also analyze, say, Kaniadakis entropy \cite{K1,K2}, which has a quite different form. For a Schwarzschild black hole of mass $M$, Kaniadakis entropy takes the form \cite{2109.09181}:
\begin{equation}
S_K=\frac{1}{K}\sinh{(4\pi K M^2)},
\end{equation}
where $K<1$ is the Kaniadakis parameter, usually assumed to be nonnegative. If we assume that $K$ runs with energy scale, then even setting $M^*$ as a solar mass gives a very small value for $K$, namely $K \sim O(10^{-76})$. The much smaller value of $K$ compared to $\delta$ is not surprising since the small $K$ expansion of $S_K$ is
\begin{equation}
S_K = \frac{A}{4} + \frac{K^2}{6}\left(\frac{A}{4}\right)^3+ O(K^4).
\end{equation}
This means that the correction term is cubic in $A$ instead of $A\ln A$, so that if we want large black holes to satisfy the standard area law, $K$ needs to suppress a bigger term -- hence it needs to be much smaller. If we take $M^*=10^2$ however, $K$ can be much larger: $K \sim O(10^{-4})$. Despite the fact that both Kaniadakis entropy have quite different motivation from Barrow's, the possibility that it can run with energy or time scale has also been discussed in the literature \cite{luciano, 2307.04027}. Indeed, in \cite{2307.04027} the authors estimated that during inflation $K \sim O(10^{-12})$, and argued for the case of a running $K$ -- with $K\sim O(1)$ in  the Planckian regime -- based on the fact that other cosmological constraints from late time typically gives $K \sim O(10^{-123})$ to $O(10^{-125})$ \cite{2111.00558,2112.04615}, while Baryogenesis constraint yields $K \sim O(10^{-83})$ \cite{luciano}. Our results are consistent with this picture.
In conclusion, we see that thermodynamical argument can be useful to put theoretical constraints on entropy modifications. 

\section{Acknowledgments}
Jiayi Xia thanks Guoyang Fu, Shulan Li, and Hengxin L\"u for useful discussions. The authors thank one of the anonymous reviewers for suggesting some improvements that lead to a further clarification on the choice $\delta \sim 1/A \ln A$ or $\delta \sim 1/A$.


\begin{thebibliography}{99}

\bibitem{2004.09444}
J. D. Barrow, ``The Area of a Rough Black Hole'',  {\hypersetup{urlcolor=vividviolet}\href{https://www.sciencedirect.com/science/article/pii/S0370269320304469?via\%3Dihub}{Phys. Lett. B \textbf{808} (2020) 135643}}, \href{https://arxiv.org/abs/2004.09444}{[arXiv:2004.09444 [gr-qc]]}.

\bibitem{2205.09311}
S. Di Gennaro, Y. C. Ong, ``Sign Switching Dark Energy from a Running Barrow Entropy'', {\hypersetup{urlcolor=vividviolet}\href{https://www.mdpi.com/2218-1997/8/10/541}{Universe \textbf{2022}, 8(10), 541}}, \href{https://arxiv.org/abs/2205.09311}{[arXiv:2205.09311 [gr-qc]]}.

\bibitem{2203.12010}
G. G. Luciano, E. N. Saridakis, ``Baryon Asymmetry From Barrow Entropy: Theoretical Predictions and Observational Constraints'',  {\hypersetup{urlcolor=vividviolet}\href{https://link.springer.com/article/10.1140/epjc/s10052-022-10530-7}{Eur. Phys. J. C \textbf{82} (2022) 6, 558}}, \href{https://arxiv.org/abs/2203.12010}{[arXiv:2203.12010 [gr-qc]]}.

\bibitem{2205.04138}
B. Farsi, A. Sheykhi, ``Growth of Perturbations in Tsallis and Barrow Cosmology'',  {\hypersetup{urlcolor=vividviolet}\href{https://link.springer.com/article/10.1140/epjc/s10052-022-11044-y}{Eur .Phys. J. C \textbf{82} (2022) 1111}}, \href{https://arxiv.org/abs/2205.04138v1}{[arXiv:2205.04138 [gr-qc]]}.


\bibitem{2112.10159}
S. Nojiri, S. D. Odintsov, T. Paul, ``Barrow Entropic Dark Energy: A Member of Generalized Holographic Dark Energy Family'', {\hypersetup{urlcolor=vividviolet}\href{https://www.sciencedirect.com/science/article/pii/S037026932100784X?via\%3Dihub}{Phys. Lett. B \textbf{825} (2022) 136844}}, \href{https://arxiv.org/abs/2112.10159}{[arXiv:2112.10159 [gr-qc]]}.

\bibitem{tsallis}
C. Tsallis, ``Possible Generalization of Boltzmann-Gibbs Statistics'', {\hypersetup{urlcolor=vividviolet}\href{https://link.springer.com/article/10.1007/BF01016429}{J. Statist. Phys. \textbf{52} (1988) 479}}.

\bibitem{2109.05315}
S. Nojiri, S. D. Odintsov, V. Faraoni, ``Area-Law Versus R\'enyi and Tsallis Black Hole Entropies'', {\hypersetup{urlcolor=vividviolet}\href{https://journals.aps.org/prd/abstract/10.1103/PhysRevD.104.084030}{Phys. Rev. D \textbf{104} (2021) 8, 084030}}, \href{https://arxiv.org/abs/2109.05315}{[arXiv:2109.05315 [gr-qc]]}.


\bibitem{0106080}
R. J. Adler, P. Chen, D. I. Santiago, ``The Generalized Uncertainty Principle and Black Hole Remnants'', {\hypersetup{urlcolor=vividviolet}\href{https://link.springer.com/article/10.1023/A:1015281430411}{Gen. Rel. Grav. \textbf{33} (2001) 2101}}, \href{https://arxiv.org/abs/gr-qc/0106080}{[arXiv:gr-qc/0106080]}.

\bibitem{MSalehi}
H. Mohammadi, A. Salehi, ``Friedmann Equations With the Generalized Logarithmic Modification of Barrow Entropy and Tsallis Entropy'', {\hypersetup{urlcolor=vividviolet}\href{https://www.sciencedirect.com/science/article/pii/S0370269323001284}{Phys . Lett. B \textbf{839} 137794}}.

\bibitem{2006.09385}
B. McInnes, Y. C. Ong, ``Event Horizon Wrinklification'', {\hypersetup{urlcolor=vividviolet}\href{https://iopscience.iop.org/article/10.1088/1361-6382/abce45}{Class. Quant. Grav. \textbf{38} (2021) 3, 034002}}, \href{https://arxiv.org/abs/2006.09385}{[arXiv:2006.09385 [gr-qc]]}.

\bibitem{1901.01613}
R. Vilela-Mendes, ``Commutative or Noncommutative Spacetime? Two Length Scales of Noncommutativity'', {\hypersetup{urlcolor=vividviolet}\href{https://journals.aps.org/prd/abstract/10.1103/PhysRevD.99.123006}{Phys. Rev. D \textbf{99} (2019) 12, 123006}}, \href{https://arxiv.org/abs/1901.01613}{[arXiv:1901.01613 [hep-th]]}.

\bibitem{2005.10302}
F. K. Anagnostopoulos, S. Basilakos, E. N. Saridakis, ``Observational Constraints on Barrow Holographic Dark Energy'', {\hypersetup{urlcolor=vividviolet}\href{https://link.springer.com/article/10.1140/epjc/s10052-020-8360-5}{Eur. Phys. J. C \textbf{80} (2020) 9, 826}}, \href{https://arxiv.org/abs/2005.10302}{[arXiv:2005.10302 [gr-qc]]}.

\bibitem{2010.00986}
J. D. Barrow, S. Basilakos, E. N. Saridakis, ``Big Bang Nucleosynthesis Constraints on Barrow Entropy'', 	{\hypersetup{urlcolor=vividviolet}\href{https://www.sciencedirect.com/science/article/pii/S0370269321000745?via\%3Dihub}{Phys. Lett. B \textbf{815} (2021) 136134}}, \href{https://arxiv.org/abs/2010.00986}{[arXiv:2010.00986 [gr-qc]]}. 

\bibitem{2108.10998}
G. Leon, J. Maga\~{n}a, A. Hern\`{a}ndez-Almada, M. A. Garc\'{i}a-Aspeitia, Tom\'{a}s Verdugo, V. Motta, ``Barrow Entropy Cosmology: An Observational Approach With a Hint of Stability Analysis'',
{\hypersetup{urlcolor=vividviolet}\href{https://iopscience.iop.org/article/10.1088/1475-7516/2021/12/032}{JCAP \textbf{12} (2021) 12, 032}}, \href{https://arxiv.org/abs/2108.10998}{[arXiv:2108.10998 [astro-ph.CO]]}.

\bibitem{K1}
G. Kaniadakis, ``Statistical Mechanics in the Context of Special Relativity'', {\hypersetup{urlcolor=vividviolet}\href{https://journals.aps.org/pre/abstract/10.1103/PhysRevE.66.056125}{Phys. Rev. E \textbf{66} (2002) 056125}}, \href{https://arxiv.org/abs/cond-mat/0210467}{[arXiv:cond-mat/0210467 [cond-mat.stat-mech]}.


\bibitem{K2}
G. Kaniadakis, ```Statistical Mechanics in the Context of Special Relativity II'', {\hypersetup{urlcolor=vividviolet}\href{https://journals.aps.org/pre/abstract/10.1103/PhysRevE.72.036108}{Phys. Rev. E \textbf{72} (2005) 036108}}, \href{https://arxiv.org/abs/cond-mat/0507311}{[arXiv:cond-mat/0507311 [cond-mat.stat-mech]}.

\bibitem{2109.09181}
N. Drepanou, A. Lymperis, E. N. Saridakis, K. Yesmakhanova, ``Kaniadakis Holographic Dark Energy and Cosmology'', {\hypersetup{urlcolor=vividviolet}\href{https://link.springer.com/article/10.1140/epjc/s10052-022-10415-9}{Eur. Phys. J. C \textbf{82} (2022) 5, 449}}, \href{https://arxiv.org/abs/2109.09181}{[arXiv:2109.09181 [gr-qc]]}.

\bibitem{luciano}
G. G. Luciano, ``Modified Friedmann Equations From Kaniadakis Entropy and Cosmological Implications on Baryogenesis And ${}^7$Li-Abundance'', {\hypersetup{urlcolor=vividviolet}\href{https://link.springer.com/article/10.1140/epjc/s10052-022-10285-1}{Eur. Phys. J. C \textbf{82} (2022) 314}}.


\bibitem{2307.04027}
G. Lambiase, G. G. Luciano, A. Sheykhi, ``Slow-Roll Inflation and Growth of Perturbations in Kaniadakis Modification of Friedmann Cosmology'', {\hypersetup{urlcolor=vividviolet}\href{https://link.springer.com/article/10.1140/epjc/s10052-023-12112-7}{Eur. Phys. J. C \textbf{83} (2023) 10, 936}}, \href{https://arxiv.org/abs/2307.04027}{[arXiv:2307.04027 [gr-qc]]}.


\bibitem{2111.00558}
A. Hernández-Almada, G. Leon, J. Magaña, M. A. García-Aspeitia, V. Motta, E. N. Saridakis, K. Yesmakhanova, ``Kaniadakis Holographic Dark Energy: Observational Constraints and Global Dynamics'', {\hypersetup{urlcolor=vividviolet}\href{https://academic.oup.com/mnras/article/511/3/4147/6517107}{Mon. Not. Roy. Astron. Soc. \textbf{511} (2022) 3, 4147}}, \href{https://arxiv.org/abs/2111.00558}{[arXiv:2111.00558 [astro-ph.CO]]}.

\bibitem{2112.04615}
A. Hernández-Almada, G. Leon, J. Magaña, M. A. García-Aspeitia, V. Motta, E. N. Saridakis, Kuralay Yesmakhanova, Alfredo D. Millano, ``Observational Constraints and Dynamical Analysis of Kaniadakis Horizon-Entropy Cosmology'', {\hypersetup{urlcolor=vividviolet}\href{https://academic.oup.com/mnras/article/512/4/5122/6553853}{Mon. Not. Roy. Astron. Soc. \textbf{512} (2022) 4, 5122}}, \href{https://arxiv.org/abs/2112.04615}{[arXiv:2112.04615 [astro-ph.CO]]}.


\end{thebibliography}
\end{document}